\documentclass[nonacm, sigconf,screen,10pt]{acmart}
\usepackage{multirow}
\usepackage{makecell} 
\usepackage[T1]{fontenc}

\newcommand{\sref}[1]{\S\ref{#1}}
\newcommand{\vheading}[1]{\vspace{0.05in}\noindent\textbf{#1}}

\renewcommand\footnotetextcopyrightpermission[1]{}

\begin{document}

\title[GDPR Anti-Patterns]{\Huge GDPR Anti-Patterns: How Design and Operation of\\[0.1cm] Modern Cloud-scale Systems Conflict with GDPR}

\author{Supreeth Shastri}
\affiliation{%
  \institution{\it Computer Science\\ University of Texas at Austin}
}

\author{Melissa Wasserman}
\affiliation{%
  \institution{\it School of Law\\ University of Texas at Austin}
}

\author{Vijay Chidambaram}
\affiliation{%
  \institution{\it Computer Science\\ University of Texas at Austin}
\vspace{6mm}
}
\email{ }
\renewcommand{\shortauthors}{Shastri, Wasserman, and Chidambaram}

\begin{abstract}
In recent years, our society is being plagued by unprecedented levels of privacy and security breaches. To rein in this trend, the European Union, in 2018, introduced a comprehensive legislation called the General Data Protection Regulation (GDPR). In this article, we review GDPR from a {\color{black} systems perspective, and identify how the design and operation of modern cloud-scale systems conflict with this regulation. We illustrate these conflicts via six} \emph{GDPR anti-patterns}: storing data without a clear timeline for deletion; reusing data indiscriminately; creating walled gardens and black markets; risk-agnostic data processing; hiding data breaches; making unexplainable decisions. Our findings reveal deep-rooted tussle between GDPR requirements and {\color{black} how cloud-scale systems that process personal data have evolved in the modern era. While it is imperative to avoid these anti-patterns,} we believe that achieving compliance requires comprehensive, grounds up solutions; anything short would amount to \emph{fixing a leaky faucet in a sinking ship}.

\end{abstract}


\maketitle

\section{Introduction}


The General Data Protection Regulation (GDPR) \cite{gdpr-regulation} is a European privacy law introduced to offer new rights and protections to people concerning their personal data. While at-scale monetization of personal data has existed since the dot-com era, a systemic disregard for privacy and protection of personal data is a recent phenomenon. For example, in 2017, we learnt about Equifax's negligence~\cite{equifax} in following the security protocols, which exposed the financial records of 145 million people; Yahoo!'s delayed confession~\cite{yahoo-breach} that three years ago, a theft had exposed all 3 billion of its user records; Facebook's admission~\cite{cambridge-analytica} that their APIs allowed illegal harvesting of user data, which in turn influenced the U.S. and U.K. democratic processes. 

Thus, GDPR was enacted to prevent a widespread and systemic abuse of personal data. At its core, GDPR declares the privacy and protection of personal data as a fundamental right. Accordingly, it grants users new rights, and assigns companies that collect personal data, new responsibilities. Any company dealing with the personal data of European customers is legally bound to comply with all the regulations of GDPR, or risk facing hefty financial penalties. For example, in January 2019, Google was fined~\cite{google-purpose-bundling} \texteuro50M for lacking customer's consent in personalizing advertisements across their different services.

{\color{black} In this work, we investigate the challenges that modern cloud-scale systems face in complying with GDPR. Specifically, we focus on the design principles and operational practices of these systems that conflict with the requirements of GDPR. To capture this tussle, we introduce the notion of \emph{GDPR anti-patterns}. In contrast to outright bad behavior, say storing customer passwords in plaintext, GDPR anti-patterns are those practices that serve their originally intended purpose well but violate the norms of GDPR. For example, given the commercial value of personal data, modern systems naturally evolved to store them without a clear timeline for deletion, and to reuse them across various applications. While these practices help the system be more reliable and affordable, they violate the storage- and purpose limitations of GDPR. 

Building on our work analyzing GDPR from a systems perspective~\cite{gdpr-sins, gdpr-storage}, we identify six GDPR anti-patterns that are widely present in the real world. These include storing personal data without a timeline for deletion; reusing personal data indiscriminately; creating black markets for personal data; risk-agnostic data processing; hiding data breaches; making unexplainable decisions. These anti-patterns highlight how the traditional system design goals of optimizing for performance, cost, and reliability sit at odds with GDPR's goal of data protection by design and by default. While eliminating these anti-patterns is not enough to achieve overall compliance under GDPR, ignoring these will definitely violate its intents.}

We structure the rest of this article as follows. First, we provide a brief primer on GDPR (\sref{sec-gdpr}). Next, we describe the six GDPR anti-patterns, discussing how they came to be, reviewing the conflicting regulations, and chronicling their real-world implications (\sref{sec-design}). Finally, we ruminate on the challenges and opportunities for system designers as societies embrace privacy regulations (\sref{sec-discussion}).

\begin{figure*}[t]
\centering
\includegraphics[width=0.95\textwidth]{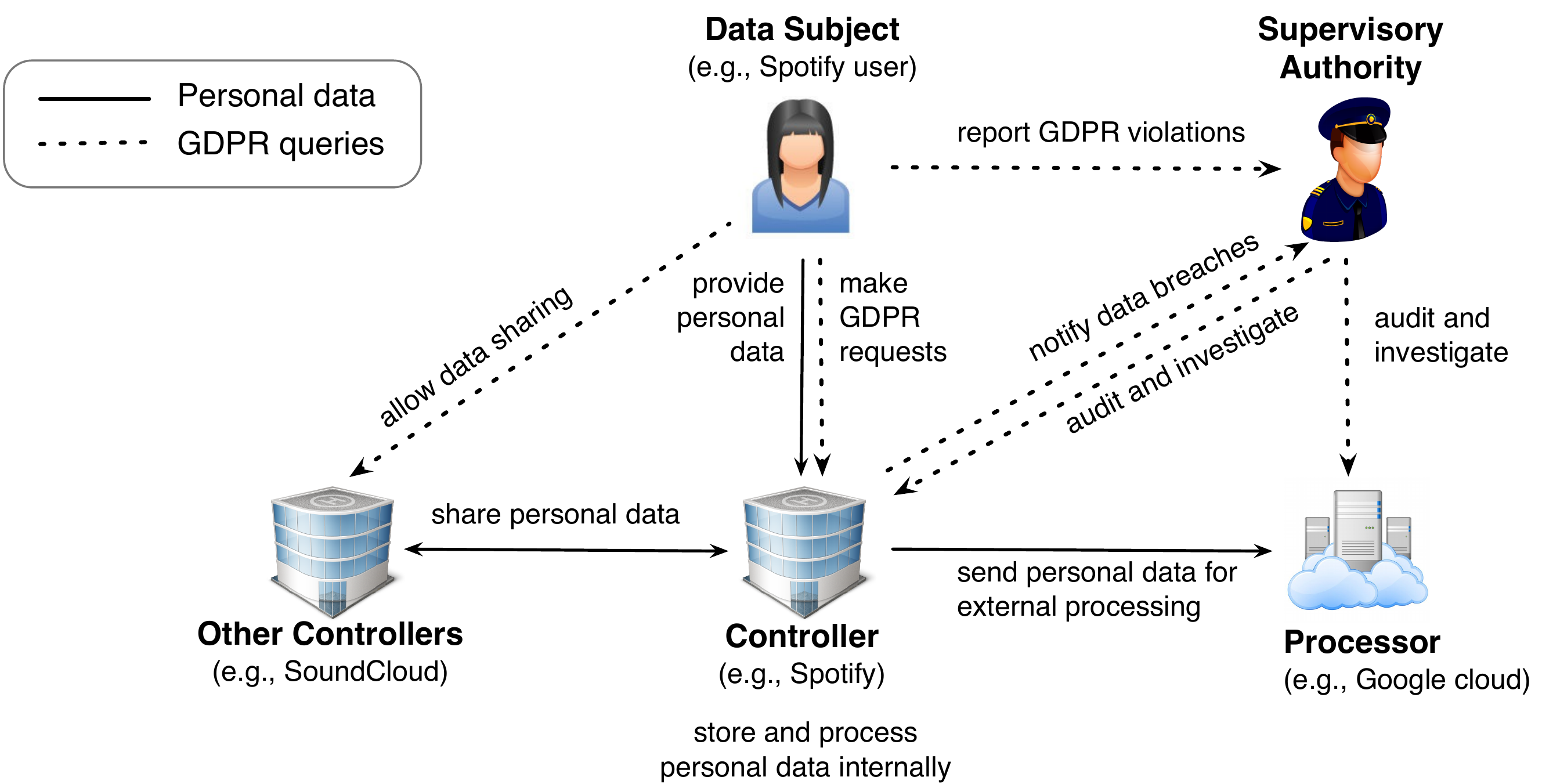}
\caption{\emph{Flow of personal data and GDPR queries between the four GDPR entities: data subjects, data controllers, data processors, and regulators.}}
\vspace{0.4cm}
\label{fig:gdpr-systems-view}
\end{figure*}

\section{GDPR}
\label{sec-gdpr}

On May 25th 2018, the European parliament adopted the General Data Protection Regulation~\cite{gdpr-regulation}. In contrast with targeted privacy regulations like HIPAA~\cite{hipaa} and FERPA~\cite{ferpa}, GDPR takes a comprehensive view by defining \emph{personal data} to be any information relating to an identifiable natural person. GDPR defines three entities that interact with personal data: (i) \emph{data subject}, the person whose personal data is collected, (ii) \emph{data controller}, the entity that collects and uses personal data, and (iii) \emph{data processor}, the entity that processes personal data on behalf of a data controller. Then, GDPR designates supervisory authorities (one per EU country) to oversee that the rights and responsibilities of GDPR are complied with.

Figure-\ref{fig:gdpr-systems-view} represents how GDPR entities interact with each other in collecting, storing, processing, securing, and sharing personal data. Consider the music streaming company Spotify collecting its customer's listening history, and then using Google cloud's services to identify new recommendations for customers. In this scenario, Spotify is the data controller and Google Cloud is the data processor. Spotify could also engage with other data controllers, say SoundCloud to gather additional personal data of their customers. 

To ensure privacy and protection of personal data in such ecosystems, GDPR grants new rights to customers and assigns responsibilities to controllers and processors. Now, any person can request a controller to grant access to all their personal data, to rectify errors, to request deletion, to object to their data being used for specific purposes, and to port their data to third parties. On the other hand, the controller is required to obtain people's consent before using their personal data, to notify them of data breaches within 72 hours of finding out, to design systems that are secure by design and by default, and to maintain records of activities performed on personal data. For controllers failing to comply with these rights and responsibilities, GDPR regulators could levy penalties of up to \texteuro20M or 4\% of their annual global revenue, whichever is higher. 

\vheading{Structure}.
GDPR is organized as 99 \emph{articles} that describe its legal requirements, and 173 \emph{recitals} that provide additional context and clarifications to these articles. The first 11 articles layout the principles of data privacy; articles 12-23 establish the rights of the people; then articles 24-50 mandate the responsibilities of the data controllers and processors; the next 26 articles describe the role and tasks of supervisory authorities; and the remainder of the articles cover liabilities, penalties and specific situations. We expand on the relevant articles in \sref{sec-design}.

\begin{table*}[t]
\makebox[1\textwidth][c]{
\begin{minipage}[b]{1\textwidth}
\centering
{\renewcommand{\arraystretch}{1.15}
\begin{tabular}{@{}l l l@{}}
\toprule[1.2pt]
{\bf Anti-Pattern} & {\bf Real-world Examples} & {\bf Governing GDPR articles} \\
\toprule[1.2pt]
\multirow{2}{*}{\makecell[l]{Storing data without a\\ clear timeline for deletion}} & \multirow{2}{*}{\makecell[l]{Search engines before Right-\\ to-be-forgotten (circa 2014)}} & \textsc{5(1e). Storage limitation} \\
& & \textsc{17. Right to be forgotten} \\ \hline
\multirow{3}{*}{Reusing data indiscriminately} & \multirow{3}{*}{\makecell[l]{Facebook collecting phone\\ numbers for 2FA and then\\ using it for ads and marketing}} & \textsc{5(1b). Purpose limitation} \\
& & \textsc{6. Lawfulness of processing} \\
& & \textsc{21. Right to object} \\ \hline
\multirow{2}{*}{Creating black markets} & \multirow{2}{*}{\makecell[l]{Illegal data harvesting by\\ programmatic ad exchanges}} & \textsc{14. Information to be provided[...]} \\
& & \textsc{20. Right to data portability}\\ \hline
\multirow{2}{*}{Risk-agnostic data processing} & \multirow{2}{*}{\makecell[l]{Strava global heatmap that\\ revealed classified military bases}} & \textsc{35. Data protection impact assessment} \\
& & \textsc{36. Prior consultation}\\ \hline
\multirow{2}{*}{Hiding data breaches} & \multirow{2}{*}{\makecell[l]{Uber paying off hackers to hide\\ their 2016 data breach}} & \textsc{5. Principles relating to processing} \\
& & \textsc{33. Notification of personal data breach}\\ \hline
\multirow{2}{*}{Making unexplainable decisions} & \multirow{2}{*}{\makecell[l]{Using software like COMPASS\\ in courts to predict recidivism}} & \textsc{15. Right of access by the data subject} \\
& & \textsc{22. Automated individual decisionmaking}\\
\bottomrule[1.2pt]
\end{tabular}
}
\end{minipage}}
\vspace{4mm}
\caption{\emph {GDPR anti-patterns, their real-world examples, and the GDPR articles that prohibit such behavior.}}
\label{tbl:gdpr-anti-patterns}
\end{table*}

\vheading{Impact}. 
Compliance with GDPR has been a challenge for many companies that collect personal data. A number of companies like Instapaper, Klout, and Unroll.me terminated their services in Europe to avoid the hassles of compliance. Few other businesses made temporary modifications. For example, media site \emph{USA Today} turned off all advertisements~\cite{usa-today-no-ads}, whereas \emph{the New York Times} stopped serving personalized ads~\cite{nytimes-no-ads}. While most organizations are working towards compliance, Gartner reports~\cite{gartner-prediction} that less than 50\% of the companies affected by GDPR were compliant by the end of 2018. This challenge is further exacerbated by the performance impact that GDPR-compliance imposes on current systems~\cite{gdpr-storage}.

In contrast, people have been enthusiastically exercising their newfound rights, and not been shy to report any shortcomings. In fact, the EU data protection board reports~\cite{gdpr-in-numbers} having received 94622 complaints from individuals and organizations in the first 9 months of GDPR. Surprisingly, even the companies have been forthcoming in reporting their security failures and data breaches, with 64684 breach notifications sent to regulators in the same 9 month period. {\color{black} In 2019, several companies have been levied hefty penalties for GDPR violations: \texteuro50 million for Google~\cite{google-purpose-bundling}, \textsterling99M for Marriott International~\cite{marriott-penalty}, and \textsterling183M for British Airways~\cite{british-airways-penalty}}.

\section{GDPR Anti-patterns}
\label{sec-design}

{\color{black}The notion of anti-patterns was first introduced~\cite{koenig-antipattern} by Andrew Koenig to characterize patterns of software design and behavior that superficially look like a good solution but ends up being counterproductive in reality. An example of this is performing premature optimizations in software systems. Extending this notion, we define the term \emph{GDPR anti-patterns} to refer to system designs and operational practices, which are effective in their own context but violate the rights and regulations of GDPR. Naturally, our definition does not include design choices that are bad in their own right, say storing customer passwords in plaintext, though they also violate GDPR norms. In this section, we catalog six GDPR anti-patterns, detailing how they came to be, which regulations they violate, and their implications in the real-world. 

\vheading{Genesis.} GDPR anti-patterns presented here have evolved from the practices and design considerations of the post dot-com era (circa 2000). These modern cloud-scale systems could be characterized by their quest for unprecedented scalability, reliability, and affordability. For example, Google operates 8 global-scale applications at 99.99\% uptime with each of them supporting more than 1 billion users. Similarly, Amazon's cloud computing infrastructure provides on-demand access to inexpensive computing to over 1 million users in 190 countries, all the while guaranteeing four nines of availability. This exclusive focus on performance, cost-efficiency, reliability, and scalability has resulted in pushing security and privacy to a backseat. 

While our GDPR analysis recognizes six anti-patterns, this list is not comprehensive. There are many other unsavory practices that would not stand the regulator scrutiny. For example, the design and operation of consent-free behavioral tracking~\cite{behavioral-ads}. Our goal here is to highlight how some of the design principles, architectural components, and operational practices of the modern cloud-scale systems conflict with the rights and responsibilities laid out in GDPR.} We present six such anti-patterns below, and summarize them in Table-\ref{tbl:gdpr-anti-patterns}.

\subsection{Storing Data Without a Clear Timeline for Deletion} 
Computing systems have always relied on insights derived from data. However, this dependence is reaching new heights, especially in this decade, with widespread adoption of machine learning and big data analytics in system design. Data has been compared to oil, electricity, gold, and even bacon~\cite{data-is-bacon}. Naturally, technology companies evolved to not only collect user data aggressively but also to preserve them forever. However, GDPR mandates that no data lives without a clear timeline for deletion.

\begin{quote} 
\textsc{Article 17: Right to be forgotten.} \textsl{``(1) The data subject shall have the right to obtain from the controller the erasure of personal data without undue delay [...]''}
\end{quote}

\begin{quote}
\textsl{Article 13: \textsc{Information to be provided where personal data are collected from the data subject.} ``(2)(a) [...] the controller shall provide the period for which the personal data will be stored, or the criteria used to determine that period;''}
\end{quote}

\begin{quote} 
\textsl {Article 5(1)(e): \textsc{Storage limitation.} ``[...] kept for no longer than is necessary for the purposes for which the personal data are processed [...]''}
\end{quote}

GDPR grants users an unconditional right, via article 17, to request their personal data be removed from everywhere in the system within a reasonable time. In conjunction with this, articles 5 and 13 lay out additional responsibilities for the data controller: (i) at the point of collection, users should be informed the time period for which their personal data would be stored, and (ii) if the personal data is no longer necessary for the purpose for which it was collected, then it should be deleted. These simply mean that all personal data should have a time-to-live (TTL) that users are aware of, and that controllers honor. However, this restriction does not apply to archiving in the public interest, or for scientific or historical research purposes. 

\vheading{Deletion in the real-world}. While conceptually clear, a timely and guaranteed removal of data is challenging in practice. For example, Google cloud describes the deletion of customer data as an iterative process~\cite{google-deletion} that could take up to 180 days to fully complete. This is because, for performance, reliability, and scalability reasons, parts of data gets replicated in various storage subsystems like memory, cache, disks, tapes, and network storage; multiple copies of data is saved in redundant backups and geographically distributed datacenters. Such practices not only delay the timeliness of deletions but also make it harder to guarantee it. 

\subsection{Reusing Data Indiscriminately}

While designing software systems, a purpose is typically associated with programs and models, whereas data is viewed as a helper resource that serves these high-level entities in accomplishing their goals. This portrayal of data as an inert entity allows it to be used freely and fungibly across various systems. For example, this has enabled organizations like Google and Amazon to collect user data once, and use it to personalize their experiences across several services. However, GDPR regulations prohibit this practice. 
 
\begin{quote} 
\textsc {Article 5(1)(b): Purpose limitation.} \textsl{``Personal data shall be collected for specified, explicit and legitimate purposes and not further processed in a manner that is incompatible with those purposes [...]''}
\end{quote}

\begin{quote} 
\textsc {Article 6: Lawfulness of processing.} \textsl{``(1)(a) Processing shall be lawful only if [...] the data subject has given consent to the processing of his or her personal data for one or more specific purposes.''}
\end{quote}

\begin{quote} 
\textsc {Article 21: Right to object.} \textsl{``(1) The data subject shall have the right to object at any time to processing of personal data concerning him or her [...]''}
\end{quote}

The first two articles establish that personal data could only be collected for specific purposes and not be used for anything else. Then, article 21 grants users a right to object, at any time, to their personal data being used for any purpose including marketing, scientific research, historical archiving, or profiling. Together, these articles require each personal data item to have its own blacklisted and whitelisted purposes that could be changed over time. 

\vheading{Purpose in the real-world}. The impact of the purpose requirement has been swift and consequential. For example, in January 2019, the French data protection commission~\cite{google-purpose-bundling} fined Google \texteuro50M for not having a legal basis for their ads personalization. Specifically, the ruling said that the user consent obtained by Google was not ``specific'' enough, and the personal data thus obtained should not have been used across 20 services.

\subsection{Walled Gardens and Black Markets}
As we are in the early days of large-scale commoditization of personal data, the norms for acquiring, sharing, and reselling them are not yet well established. This has led to uncertainties for people and a tussle for control over data amongst controllers. People are concerned about vendor lock-ins, and about a lack of visibility once their data is resold or shared in the secondary markets. Organizations have responded to this by setting up walled gardens, and making secondary markets more opaque. However, GDPR dismantles such practices.

\begin{quote} 
\textsc {Article 20: Right to data portability.} \textsl{``(1) The data subject shall have the right to receive the personal data concerning him or her, which he or she has provided to a controller. (2) [...] the right to have the personal data transmitted directly from one controller to another.''}
\end{quote}

\begin{quote} 
\textsc {Article 14: Information to be provided where personal data have not been obtained from the data subject.} \textsl{``(1) (c) the purposes of the processing [...], (e) the recipients [...], (2) (a) the period for which the personal data will be stored [...], (f) from which source the personal data originate [...]. (3) The controller shall provide the information at the latest within one month.''}
\end{quote}

With article 20, people have a right to request for all the personal data that a controller has collected directly from them. Not only that, they could also ask the controller to directly transmit all such personal data to a different controller. While that tackles the vendor lock-ins, article 14 regulates the behavior in secondary markets. It requires that anyone indirectly procuring personal data must inform the users, within a month, about (i) how they acquired it, (ii) how long would they be stored, (iii) what purpose would they be used for, and (iv) who they intend to share it with. The \emph{data trail} set up by this regulation should bring the control and clarity back to the people.

\vheading{Data movement in the real-world.} When GDPR went live, a large number of companies rolled out~\cite{gdpr-data-download} data download tools for EU users. For example, Google Takeout~\cite{google-takeout} lets users not only access all their personal data in their system but also port data directly to external services. However, the impact has been less savory for programmatic ad exchanges~\cite{gdpr-programmatic-ad-buy} in Europe, many of which had to shut down. This was primarily due to Google and Facebook restricting access to their platforms for those ad exchanges, which could not verify the legality of the personal data they possessed.

\subsection{Risk-Agnostic Data Processing}

Modern technology companies face the challenge of creating and managing increasingly complex software systems in an environment that demands rapid innovation. This has led to a practice, especially in the Internet-era companies, of prioritizing speed over correctness; and to a belief~\cite{move-fast-break-things} that \textit{unless you are breaking stuff, you are not moving fast enough}~\cite{zuckerberg-quote}. However, GDPR explicitly restricts this approach when dealing with personal data. 

\begin{quote} 
\textsc {Article 35: Data protection impact assessment.} \textsl{``(1) Where processing, in particular using new technologies, is likely to result in a high risk to the rights and freedoms of natural persons, the controller shall, prior to the processing, carry out an assessment of the impact of the envisaged processing operations on the protection of personal data.''}
\end{quote}

\begin{quote} 
\textsc {Article 36: Prior consultation.} \textsl{``(1) The controller shall consult the supervisory authority prior to processing where [...] that would result in a high risk in the absence of measures taken by the controller to mitigate the risk.''}
\end{quote}

GDPR establishes, via articles 35 and 36, two levels of checks for introducing new technologies and for modifying existing systems, if they process large amounts of personal data. The first level is internal to the controller, where an impact assessment must analyze the nature and scope of the risks, and then propose the safeguards needed to mitigate them. Next, if the risks are systemic in nature or concern common platforms, either internal and external, the data protection officer must consult with the supervisory authority prior to any processing.

\vheading{Fast and broken in the real-world}. Facebook, despite having moved away from the aforementioned motto, has continued to be plagued by it. In 2018, it revealed two major breaches: first, that their APIs allowed Cambridge Analytica to illicitly harvest~\cite{cambridge-analytica} personal data from 87M users, and then their new \emph{View As} feature was exploited~\cite{facebook-view-as-leak} to gain control over 50M user accounts. However, this practice of prioritizing speed over security is not limited to one organization. For example, in Nov 2017, fitness app Strava released an athlete motivation tool called global heatmap~\cite{strava-heatmap} that visualized athletic activities of worldwide users. However, within months, these maps were used to identify undisclosed military bases and covert security operations~\cite{strava-military-leak}, jeopardizing missions and lives of soldiers. 

\subsection{Hiding Data Breaches}

The notion that one is \emph{innocent until proven guilty} predates all computer systems. As a legal principle, it dates back to 6th century Roman empire~\cite{roman-law}, where it was codified that \emph{proof lies on him who asserts, not on him who denies}. Thus, in the event of a data breach or a privacy violation, organizations typically claim innocence and ignorance, and seek to be absolved of their responsibilities. However, GDPR makes such presumption conditional on the controller proactively implementing risk-appropriate security measures (i.e., accountability), and notifying breaches in a timely fashion (i.e., transparency).

\begin{quote} 
\textsc {Article 5: Principles relating to processing.} \textsl{``(1) Personal data shall be processed with [...] lawfulness, fairness and transparency; [...] purpose limitation; [...] data minimisation; [...] accuracy; [...] storage limitation; [...] integrity and confidentiality. (2) The controller shall be responsible for, and be able to demonstrate compliance with (1).''}
\end{quote}

\begin{quote} 
\textsc {Article 33: Notification of a personal data breach.} \textsl{``(1) the controller shall without undue delay and not later than 72 hours after having become aware of it, notify the supervisory authority. [...] (3) The notification shall at least describe the nature of the personal breach, [...] likely consequences, and [...] measures taken to mitigate its adverse effects.''}
\end{quote}

GDPR's goal is two folds: first, to reduce the frequency and impact of data breaches, article 5 lays out several ground rules. The controllers are not only expected to adhere to these internally but also be able to demonstrate their compliance externally. Second, to bring transparency in handling data breaches, articles 33 and 34 mandate a 72 hour notification window within which the controller should inform both the supervisory authority and the affected people. 

\vheading{Data breaches in the real-world.} In recent years, responses to personal data breaches have been ad hoc: while a few organizations have been forthcoming, others have chosen to refute~\cite{aadhar-refute-breach}, delay~\cite{panera-delay-breach} or hide by paying off hackers~\cite{uber-breach-payoff}. However, GDPR's impact has been swift and clear. Just in the first 8 months (May 2018 to Jan 2019), regulators have received 41,502 data breach notifications~\cite{gdpr-in-numbers}. This number is in stark contrast from the pre-GDPR era, with reports of 945 worldwide data breaches in the first half of 2018~\cite{pre-gdpr-numbers}. 

\subsection{Making Unexplainable Decisions}
Algorithmic decision-making has been successfully applied to several domains: curating media content, managing industrial operations, trading financial instruments, personalizing advertisements, and even combating fake news. Their inherent efficiency and scalability (with no human in the loop) has made them a necessity in modern system design. However, GDPR takes a cautious view of this trend.

\begin{quote} 
\textsc {Article 22: Automated individual decision-making.} \textsl{``(1) The data subject shall have the right not to be subject to a decision based solely on automated processing [...]''}
\end{quote}

\begin{quote} 
\textsc {Article 15: Right of access by the data subject.} \textsl{``(1) The data subject shall have the right to obtain from the controller [...] meaningful information about the logic involved, as well as the significance and the envisaged consequences of such processing.''}
\end{quote}

On one hand, privacy researchers from Oxford postulate~\cite{gdpr-explanation} that these two regulations, together with recital 71, establish a ``right to explanation'' and thus, human interpretability should be a design consideration for machine learning and artificial intelligence systems. However, another group at Oxford argues~\cite{gdpr-no-right-to-explanation} that GDPR falls short of mandating this right by requiring users (i) to demonstrate significant consequences, (ii) to seek explanation only after a decision has been made, and (iii) to have to opt out explicitly.

\vheading{Decision-making in the real-world.} The debate over the privacy and interpretability in automated decision-making has just begun. Starting 2016, the machine learning and intelligence community began exploring this rigorously: the workshop on Explainable AI~\cite{xAI} at IJCAI, and the workshop on Human Interpretability in Machine Learning~\cite{WHI} at ICML being two such efforts. In January 2019, privacy advocacy group NoYB has filed~\cite{noyb-streaming-service-complaints} complaints against 8 streaming services including Amazon, Apple Music, Netflix, SoundCloud, Spotify, YouTube, Flimmit and DAZN for violating article 15 requirements in their recommendation systems.

\section{Concluding Remarks}
\label{sec-discussion}

{\color{black} Achieving compliance with GDPR, while mandatory for companies working with personal data of Europeans, is not trivial. In this paper, we examine how the design, architecture, and operation of modern cloud-scale systems conflict with GDPR. Specifically, we illustrate this tussle via six \emph{GDPR anti-patterns} i.e., patterns of system design and operation, which are effective in their own context but violate the rights and regulations of GDPR. Given the importance of personal data, and the implications of misusing them, we believe that system designers should examine their systems for these anti-patterns, and work towards eliminating them with urgency.}

\vheading{Open issues}.
{\color{black} While our list of GDPR anti-patterns is a useful beginning point, it is not exhaustive. Neither have we proposed a methodology for identifying a large number of such anti-patterns, nor do we prescribe any mechanisms towards eliminating them. The six} anti-patterns highlighted here exist due to technical and economical reasons that may not entirely be in the control of individual companies. Thus, solving such deep rooted issues would likely result in significant performance overheads, slower product rollouts, and reorganization of data markets. The equilibrium point of these tussles are not yet clear. 

\vheading{Future directions}.
{\color{black} While there have been a number of recent work analyzing GDPR from privacy and legal perspectives~\cite{poly-privacy-policy, cacm-gdpr-impact, counterfactual, gdpr-purpose, gdpr-www, explainable-machines, uninformed-consent, gdpr-cookies, gdpr-formal}, the systems community is just beginning to get involved. GDPR compliance brings several interesting challenges to system design. Prominently, addressing compliance at the level of individual infrastructure components (i.e., compute, storage, and networking) versus achieving end-to-end compliance of individual regulations (i.e., implementing right-of-access in a music streaming service) will result in different tradeoffs. The former approach makes the effort more contained and thus, suits the cloud model better. Examples of this direction include GDPR-compliant Redis~\cite{gdpr-storage}, Compliance by construction~\cite{poly-compliance-by-construction}, and Data protection database~\cite{poly-datumdb}. The latter approach provides opportunities for cross-layer optimizations (e.g., avoiding access control in multiple layers). Google search's implementation~\cite{google-forgotten} of Right to be forgotten is in this direction.} 

Another challenge arises from GDPR being vague in its technical specifications (possibly to allow for technological advancements). Thus, questions like \emph{how soon after a delete request should that data be actually deleted} could be answered in several compliant ways. The idea that compliance could be a spectrum, instead of a well-defined point gives rise to interesting system tradeoffs as well as the need for benchmarks that quantify a given system's compliance behavior.  

While GDPR is the first comprehensive privacy legislation in the world, several governments are actively drafting their own privacy regulations. For instance, California'a Consumer Privacy Act (CCPA)~\cite{ccpa}, which goes into effect on Jan 1, 2020. We hope that this paper helps all the stakeholders in avoiding the pitfalls in designing and operating GDPR-compliant personal-data processing systems.

\bibliographystyle{ACM-Reference-Format}
\bibliography{paper}


\begin{thebibliography}{50}


\ifx \showCODEN    \undefined \def \showCODEN     #1{\unskip}     \fi
\ifx \showDOI      \undefined \def \showDOI       #1{#1}\fi
\ifx \showISBNx    \undefined \def \showISBNx     #1{\unskip}     \fi
\ifx \showISBNxiii \undefined \def \showISBNxiii  #1{\unskip}     \fi
\ifx \showISSN     \undefined \def \showISSN      #1{\unskip}     \fi
\ifx \showLCCN     \undefined \def \showLCCN      #1{\unskip}     \fi
\ifx \shownote     \undefined \def \shownote      #1{#1}          \fi
\ifx \showarticletitle \undefined \def \showarticletitle #1{#1}   \fi
\ifx \showURL      \undefined \def \showURL       {\relax}        \fi
\providecommand\bibfield[2]{#2}
\providecommand\bibinfo[2]{#2}
\providecommand\natexlab[1]{#1}
\providecommand\showeprint[2][]{arXiv:#2}

\bibitem[\protect\citeauthoryear{??}{fer}{1974}]%
        {ferpa}
 \bibinfo{year}{1974}\natexlab{}.
\newblock \showarticletitle{{Family Educational Rights and Privacy Act}}.
\newblock \bibinfo{journal}{\emph{Title 20 of the United States Code, Section
  1232g}} (\bibinfo{date}{Aug 21} \bibinfo{year}{1974}).
\newblock


\bibitem[\protect\citeauthoryear{??}{hip}{1996}]%
        {hipaa}
 \bibinfo{year}{1996}\natexlab{}.
\newblock \showarticletitle{{The Health Insurance Portability and
  Accountability Act}}.
\newblock \bibinfo{journal}{\emph{104th United States Congress Public Law 191}}
  (\bibinfo{date}{Aug 21} \bibinfo{year}{1996}).
\newblock


\bibitem[\protect\citeauthoryear{??}{ccp}{2018}]%
        {ccpa}
 \bibinfo{year}{2018}\natexlab{}.
\newblock \showarticletitle{{California Consumer Privacy Act}}.
\newblock \bibinfo{journal}{\emph{California Civil Code, Section 1798.100}}
  (\bibinfo{date}{Jun 28} \bibinfo{year}{2018}).
\newblock


\bibitem[\protect\citeauthoryear{??}{goo}{2018}]%
        {google-deletion}
 \bibinfo{year}{2018}\natexlab{}.
\newblock \bibinfo{title}{{Data Deletion on Google Cloud Platform}}.
\newblock
  \bibinfo{howpublished}{\url{https://cloud.google.com/security/deletion/}}.
\newblock


\bibitem[\protect\citeauthoryear{??}{goo}{2019}]%
        {google-takeout}
 \bibinfo{year}{2019}\natexlab{}.
\newblock \bibinfo{title}{Google {T}akeout}.
\newblock \bibinfo{howpublished}{\url{https://takeout.google.com/}}.
\newblock


\bibitem[\protect\citeauthoryear{Aha, Darrell, Pazzani, Reid, Sammut, and
  Stone}{Aha et~al\mbox{.}}{2017}]%
        {xAI}
\bibfield{editor}{\bibinfo{person}{David Aha}, \bibinfo{person}{Trevor
  Darrell}, \bibinfo{person}{Michael Pazzani}, \bibinfo{person}{Darryn Reid},
  \bibinfo{person}{Claude Sammut}, {and} \bibinfo{person}{Peter Stone}} (Eds.).
  \bibinfo{year}{2017}\natexlab{}.
\newblock \bibinfo{booktitle}{\emph{Workshop on Explainable Artificial
  Intelligence}}. International Joint Conference on Artificial Intelligence
  (IJCAI).
\newblock


\bibitem[\protect\citeauthoryear{Alexander}{Alexander}{2016}]%
        {data-is-bacon}
\bibfield{author}{\bibinfo{person}{Forsyth Alexander}.}
  \bibinfo{year}{2016}\natexlab{}.
\newblock \bibinfo{title}{Data is the new bacon. {I}n \emph{IBM Business
  analytics blog}}.
\newblock
  \bibinfo{howpublished}{\url{https://www.ibm.com/blogs/business-analytics/data-is-the-new-bacon/}}.
\newblock


\bibitem[\protect\citeauthoryear{Basin, Debois, and Hildebrandt}{Basin
  et~al\mbox{.}}{2018}]%
        {gdpr-purpose}
\bibfield{author}{\bibinfo{person}{David Basin}, \bibinfo{person}{S{\o}ren
  Debois}, {and} \bibinfo{person}{Thomas Hildebrandt}.}
  \bibinfo{year}{2018}\natexlab{}.
\newblock \showarticletitle{On {P}urpose and by {N}ecessity: {C}ompliance under
  the {GDPR}}. In \bibinfo{booktitle}{\emph{Financial Cryptography and Data
  Security}}.
\newblock


\bibitem[\protect\citeauthoryear{Bertram, Bursztein, Caro, Chao, Feman,
  Fleischer, Gustafsson, Hemerly, Hibbert, and Invernizzi}{Bertram
  et~al\mbox{.}}{2018}]%
        {google-forgotten}
\bibfield{author}{\bibinfo{person}{Theo Bertram}, \bibinfo{person}{Elie
  Bursztein}, \bibinfo{person}{Stephanie Caro}, \bibinfo{person}{Hubert Chao},
  \bibinfo{person}{Rutledge Feman}, \bibinfo{person}{Peter Fleischer},
  \bibinfo{person}{Albin Gustafsson}, \bibinfo{person}{Jess Hemerly},
  \bibinfo{person}{Chris Hibbert}, {and} \bibinfo{person}{Luca Invernizzi}.}
  \bibinfo{year}{2018}\natexlab{}.
\newblock \bibinfo{booktitle}{\emph{Three years of the {R}ight to be
  {F}orgotten}}.
\newblock \bibinfo{type}{Technical Report}. \bibinfo{institution}{Google Inc.}
\newblock


\bibitem[\protect\citeauthoryear{Blodget}{Blodget}{2009}]%
        {zuckerberg-quote}
\bibfield{author}{\bibinfo{person}{Henry Blodget}.}
  \bibinfo{year}{2009}\natexlab{}.
\newblock \bibinfo{title}{Mark Zuckerberg On Innovation. {I}n \emph{Business
  Insider}}.
\newblock
\newblock


\bibitem[\protect\citeauthoryear{Board}{Board}{2019}]%
        {gdpr-in-numbers}
\bibfield{author}{\bibinfo{person}{The European Data~Protection Board}.}
  \bibinfo{year}{2019}\natexlab{}.
\newblock \bibinfo{title}{{GDPR} in {N}umbers}.
\newblock
  \bibinfo{howpublished}{\url{https://ec.europa.eu/commission/sites/beta-political/files/190125_gdpr_infographics_v4.pdf}}.
\newblock


\bibitem[\protect\citeauthoryear{Buckland and Stein}{Buckland and
  Stein}{2007}]%
        {roman-law}
\bibfield{author}{\bibinfo{person}{William Buckland} {and}
  \bibinfo{person}{Peter Stein}.} \bibinfo{year}{2007}\natexlab{}.
\newblock \bibinfo{booktitle}{\emph{{A text-book of Roman law: From Augustus to
  Justinian}}}.
\newblock \bibinfo{publisher}{Cambridge University Press}.
\newblock


\bibitem[\protect\citeauthoryear{Casey, Farhangi, and Vogl}{Casey
  et~al\mbox{.}}{2019}]%
        {explainable-machines}
\bibfield{author}{\bibinfo{person}{Bryan Casey}, \bibinfo{person}{Ashkon
  Farhangi}, {and} \bibinfo{person}{Roland Vogl}.}
  \bibinfo{year}{2019}\natexlab{}.
\newblock \showarticletitle{Rethinking Explainable Machines: The GDPR's Right
  to Explanation Debate and the Rise of Algorithmic Audits in Enterprise}.
\newblock \bibinfo{journal}{\emph{Berkeley Technology Law Journal}}
  \bibinfo{volume}{34} (\bibinfo{year}{2019}), \bibinfo{pages}{143}.
\newblock


\bibitem[\protect\citeauthoryear{CNIL}{CNIL}{2019}]%
        {google-purpose-bundling}
\bibfield{author}{\bibinfo{person}{CNIL}.} \bibinfo{year}{2019}\natexlab{}.
\newblock \bibinfo{title}{The {CNIL}’s restricted committee imposes a
  financial penalty of 50 Million euros against {Google LLC}}.
\newblock
  \bibinfo{howpublished}{\url{https://www.cnil.fr/en/cnils-restricted-committee-imposes-financial-penalty-50-million-euros-against-google-llc}}.
\newblock


\bibitem[\protect\citeauthoryear{Conger}{Conger}{2018}]%
        {gdpr-data-download}
\bibfield{author}{\bibinfo{person}{Kate Conger}.}
  \bibinfo{year}{2018}\natexlab{}.
\newblock \bibinfo{title}{{How to Download Your Data With All the Fancy New
  {GDPR} Tools}. {I}n \emph{Gizmodo}}.
\newblock
  \bibinfo{howpublished}{\url{https://gizmodo.com/how-to-download-your-data-with-all-the-fancy-new-gdpr-t-1826334079}}.
\newblock


\bibitem[\protect\citeauthoryear{Davies}{Davies}{2018}]%
        {gdpr-programmatic-ad-buy}
\bibfield{author}{\bibinfo{person}{Jessica Davies}.}
  \bibinfo{year}{2018}\natexlab{}.
\newblock \bibinfo{title}{{GDPR mayhem: Programmatic ad buying plummets in
  Europe}. {I}n \emph{Digiday}}.
\newblock
  \bibinfo{howpublished}{\url{https://digiday.com/media/gdpr-mayhem-programmatic-ad-buying-plummets-europe/}}.
\newblock


\bibitem[\protect\citeauthoryear{Davies}{Davies}{2019}]%
        {nytimes-no-ads}
\bibfield{author}{\bibinfo{person}{Jessica Davies}.}
  \bibinfo{year}{2019}\natexlab{}.
\newblock \bibinfo{title}{{After GDPR, The New York Times cut off ad exchanges
  in Europe}. {I}n \emph{Digiday}}.
\newblock
  \bibinfo{howpublished}{\url{https://digiday.com/media/new-york-times-gdpr-cut-off-ad-exchanges-europe-ad-revenue/}}.
\newblock


\bibitem[\protect\citeauthoryear{Degeling, Utz, Lentzsch, Hosseini, Schaub, and
  Holz}{Degeling et~al\mbox{.}}{2019}]%
        {gdpr-cookies}
\bibfield{author}{\bibinfo{person}{Martin Degeling}, \bibinfo{person}{Christine
  Utz}, \bibinfo{person}{Christopher Lentzsch}, \bibinfo{person}{Henry
  Hosseini}, \bibinfo{person}{Florian Schaub}, {and} \bibinfo{person}{Thorsten
  Holz}.} \bibinfo{year}{2019}\natexlab{}.
\newblock \showarticletitle{We Value Your Privacy... Now Take Some Cookies:
  Measuring the GDPR's Impact on Web Privacy}. In
  \bibinfo{booktitle}{\emph{NDSS}}.
\newblock


\bibitem[\protect\citeauthoryear{Doshi}{Doshi}{2018}]%
        {aadhar-refute-breach}
\bibfield{author}{\bibinfo{person}{Vidhi Doshi}.}
  \bibinfo{year}{2018}\natexlab{}.
\newblock \bibinfo{title}{{A security breach in India has left a billion people
  at risk of identity theft}. {I}n \emph{The Washington Post}}.
\newblock
  \bibinfo{howpublished}{\url{https://www.washingtonpost.com/news/worldviews/wp/2018/01/04/a-security-breach-in-india-has-left-a-billion-people-at-risk-of-identity-theft}}.
\newblock


\bibitem[\protect\citeauthoryear{Forni and van~der Meulen}{Forni and van~der
  Meulen}{2017}]%
        {gartner-prediction}
\bibfield{author}{\bibinfo{person}{Amy~Ann Forni} {and} \bibinfo{person}{Rob
  van~der Meulen}.} \bibinfo{year}{2017}\natexlab{}.
\newblock \bibinfo{title}{Organizations Are Unprepared for the 2018 {E}uropean
  {D}ata {P}rotection {R}egulation. {I}n \emph{Gartner}}.
\newblock
\newblock


\bibitem[\protect\citeauthoryear{Goodman and Flaxman}{Goodman and
  Flaxman}{2017}]%
        {gdpr-explanation}
\bibfield{author}{\bibinfo{person}{Bryce Goodman} {and} \bibinfo{person}{Seth
  Flaxman}.} \bibinfo{year}{2017}\natexlab{}.
\newblock \showarticletitle{European {U}nion {R}egulations on {A}lgorithmic
  {D}ecision-{M}aking and a {R}ight to {E}xplanation}.
\newblock \bibinfo{journal}{\emph{AAAI AI Magazine}} \bibinfo{volume}{38},
  \bibinfo{number}{3} (\bibinfo{year}{2017}).
\newblock


\bibitem[\protect\citeauthoryear{Greengard}{Greengard}{2018}]%
        {cacm-gdpr-impact}
\bibfield{author}{\bibinfo{person}{Samuel Greengard}.}
  \bibinfo{year}{2018}\natexlab{}.
\newblock \showarticletitle{Weighing the impact of GDPR}.
\newblock \bibinfo{journal}{\emph{Commun. ACM}} \bibinfo{volume}{61},
  \bibinfo{number}{11} (\bibinfo{year}{2018}), \bibinfo{pages}{16--18}.
\newblock


\bibitem[\protect\citeauthoryear{Grothaus}{Grothaus}{2018}]%
        {panera-delay-breach}
\bibfield{author}{\bibinfo{person}{Michael Grothaus}.}
  \bibinfo{year}{2018}\natexlab{}.
\newblock \bibinfo{title}{{Panera Bread leaked millions of customers’ data}.
  {I}n \emph{Fast Company}}.
\newblock
  \bibinfo{howpublished}{\url{https://www.fastcompany.com/40553518/report-panera-bread-leaked-millions-of-customers-data}}.
\newblock


\bibitem[\protect\citeauthoryear{Haselton}{Haselton}{2017}]%
        {equifax}
\bibfield{author}{\bibinfo{person}{Todd Haselton}.}
  \bibinfo{year}{2017}\natexlab{}.
\newblock \bibinfo{title}{{Credit reporting firm Equifax says data breach could
  potentially affect 143 million US consumers.} {I}n \emph{CNBC}}.
\newblock
\newblock


\bibitem[\protect\citeauthoryear{Isaac, Benner, and Frenkel}{Isaac
  et~al\mbox{.}}{2017}]%
        {uber-breach-payoff}
\bibfield{author}{\bibinfo{person}{Mike Isaac}, \bibinfo{person}{Katie Benner},
  {and} \bibinfo{person}{Sheera Frenkel}.} \bibinfo{year}{2017}\natexlab{}.
\newblock \bibinfo{title}{{Uber Hid 2016 Breach, Paying Hackers to Delete
  Stolen Data.} {I}n \emph{The New York Times}}.
\newblock
  \bibinfo{howpublished}{\url{https://www.nytimes.com/2017/11/21/technology/uber-hack.html}}.
\newblock


\bibitem[\protect\citeauthoryear{Kammueller}{Kammueller}{2018}]%
        {gdpr-formal}
\bibfield{author}{\bibinfo{person}{Florian Kammueller}.}
  \bibinfo{year}{2018}\natexlab{}.
\newblock \showarticletitle{Formal modeling and analysis of data protection for
  gdpr compliance of IoT healthcare systems}. In \bibinfo{booktitle}{\emph{IEEE
  International Conference on Systems, Man, and Cybernetics (SMC)}}.
\newblock


\bibitem[\protect\citeauthoryear{Kim, Malioutov, and Varshney}{Kim
  et~al\mbox{.}}{2016}]%
        {WHI}
\bibfield{editor}{\bibinfo{person}{Been Kim}, \bibinfo{person}{Dmitry
  Malioutov}, {and} \bibinfo{person}{Kush Varshney}} (Eds.).
  \bibinfo{year}{2016}\natexlab{}.
\newblock \bibinfo{booktitle}{\emph{Workshop on Human Interpretability in
  Machine Learning}}. International Conference on Machine Learning (ICML).
\newblock


\bibitem[\protect\citeauthoryear{Koenig}{Koenig}{1995}]%
        {koenig-antipattern}
\bibfield{author}{\bibinfo{person}{Andrew Koenig}.}
  \bibinfo{year}{1995}\natexlab{}.
\newblock \showarticletitle{Patterns and Antipatterns}.
\newblock \bibinfo{journal}{\emph{Journal of Object-Oriented Programming}}
  \bibinfo{volume}{8}, \bibinfo{number}{1} (\bibinfo{year}{1995}),
  \bibinfo{pages}{46--48}.
\newblock


\bibitem[\protect\citeauthoryear{Kraska, Stonebraker, Brodie, Servan-Schreiber,
  and Weitzner}{Kraska et~al\mbox{.}}{2019}]%
        {poly-datumdb}
\bibfield{author}{\bibinfo{person}{Tim Kraska}, \bibinfo{person}{Michael
  Stonebraker}, \bibinfo{person}{Michael Brodie}, \bibinfo{person}{Sacha
  Servan-Schreiber}, {and} \bibinfo{person}{Daniel Weitzner}.}
  \bibinfo{year}{2019}\natexlab{}.
\newblock \showarticletitle{DATUMDB: A Data Protection Database Proposal}. In
  \bibinfo{booktitle}{\emph{Poly'19 co-located at VLDB 2019}}.
\newblock


\bibitem[\protect\citeauthoryear{Larson}{Larson}{2017}]%
        {yahoo-breach}
\bibfield{author}{\bibinfo{person}{Selena Larson}.}
  \bibinfo{year}{2017}\natexlab{}.
\newblock \bibinfo{title}{{Every single Yahoo! account was hacked - 3 billion
  in all}. {I}n \emph{CNN Business}}.
\newblock
\newblock


\bibitem[\protect\citeauthoryear{Lomas}{Lomas}{2019a}]%
        {behavioral-ads}
\bibfield{author}{\bibinfo{person}{Natasha Lomas}.}
  \bibinfo{year}{2019}\natexlab{a}.
\newblock \bibinfo{title}{{Even the IAB warned adtech risks EU privacy rules}.
  {I}n \emph{TechCrunch}}.
\newblock
  \bibinfo{howpublished}{\url{https://techcrunch.com/2019/02/21/even-the-iab-warned-adtech-risks-eu-privacy-rules/}}.
\newblock


\bibitem[\protect\citeauthoryear{Lomas}{Lomas}{2019b}]%
        {noyb-streaming-service-complaints}
\bibfield{author}{\bibinfo{person}{Natasha Lomas}.}
  \bibinfo{year}{2019}\natexlab{b}.
\newblock \bibinfo{title}{{Privacy campaigner Schrems slaps Amazon, Apple,
  Netflix, others with GDPR data access complaints.} {I}n \emph{TechCrunch}}.
\newblock
\newblock


\bibitem[\protect\citeauthoryear{Lunden}{Lunden}{2019}]%
        {british-airways-penalty}
\bibfield{author}{\bibinfo{person}{Ingrid Lunden}.}
  \bibinfo{year}{2019}\natexlab{}.
\newblock \bibinfo{title}{{UK's ICO fines British Airways a record £183M over
  GDPR breach that leaked data from 500000 users.} {I}n \emph{TechCrunch}}.
\newblock
\newblock


\bibitem[\protect\citeauthoryear{Mohan, Wasserman, and Chidambaram}{Mohan
  et~al\mbox{.}}{2019}]%
        {poly-privacy-policy}
\bibfield{author}{\bibinfo{person}{Jayashree Mohan}, \bibinfo{person}{Melissa
  Wasserman}, {and} \bibinfo{person}{Vijay Chidambaram}.}
  \bibinfo{year}{2019}\natexlab{}.
\newblock \showarticletitle{Analyzing GDPR Compliance Through the Lens of
  Privacy Policy}. In \bibinfo{booktitle}{\emph{Poly'19 co-located at VLDB
  2019}}.
\newblock


\bibitem[\protect\citeauthoryear{O'Flaherty}{O'Flaherty}{2019}]%
        {marriott-penalty}
\bibfield{author}{\bibinfo{person}{Kate O'Flaherty}.}
  \bibinfo{year}{2019}\natexlab{}.
\newblock \bibinfo{title}{{Marriott Faces £123 Million Fine For 2018 Mega
  Breach.} {I}n \emph{Forbes}}.
\newblock
\newblock


\bibitem[\protect\citeauthoryear{Quarles}{Quarles}{2018}]%
        {strava-military-leak}
\bibfield{author}{\bibinfo{person}{James Quarles}.}
  \bibinfo{year}{2018}\natexlab{}.
\newblock \bibinfo{title}{{An Update on the Global Heatmap}}.
\newblock
  \bibinfo{howpublished}{\url{https://blog.strava.com/press/a-letter-to-the-strava-community/}}.
\newblock


\bibitem[\protect\citeauthoryear{Regulation}{Regulation}{2016}]%
        {gdpr-regulation}
\bibfield{author}{\bibinfo{person}{General Data~Protection Regulation}.}
  \bibinfo{year}{2016}\natexlab{}.
\newblock \showarticletitle{Regulation ({EU}) 2016/679 of the {E}uropean
  {P}arliament and of the {C}ouncil of 27 {A}pril 2016 on the protection of
  natural persons with regard to the processing of personal data and on the
  free movement of such data, and repealing {D}irective 95/46}.
\newblock \bibinfo{journal}{\emph{Official Journal of the European Union}}
  \bibinfo{volume}{59}, \bibinfo{number}{1-88} (\bibinfo{year}{2016}).
\newblock


\bibitem[\protect\citeauthoryear{Robb}{Robb}{2017}]%
        {strava-heatmap}
\bibfield{author}{\bibinfo{person}{Drew Robb}.}
  \bibinfo{year}{2017}\natexlab{}.
\newblock \bibinfo{title}{{Building the Global Heatmap}}.
\newblock
  \bibinfo{howpublished}{\url{https://medium.com/strava-engineering/the-global-heatmap-now-6x-hotter-23fc01d301de}}.
\newblock


\bibitem[\protect\citeauthoryear{Rosen}{Rosen}{2018}]%
        {facebook-view-as-leak}
\bibfield{author}{\bibinfo{person}{Guy Rosen}.}
  \bibinfo{year}{2018}\natexlab{}.
\newblock \bibinfo{title}{{Security Update}}.
\newblock
  \bibinfo{howpublished}{\url{https://newsroom.fb.com/news/2018/09/security-update/}}.
\newblock


\bibitem[\protect\citeauthoryear{Schwarzkopf, Kohler, Kaashoek, and
  Morris}{Schwarzkopf et~al\mbox{.}}{2019}]%
        {poly-compliance-by-construction}
\bibfield{author}{\bibinfo{person}{Malte Schwarzkopf}, \bibinfo{person}{Eddie
  Kohler}, \bibinfo{person}{Frans Kaashoek}, {and} \bibinfo{person}{Robert
  Morris}.} \bibinfo{year}{2019}\natexlab{}.
\newblock \showarticletitle{GDPR Compliance by Construction}. In
  \bibinfo{booktitle}{\emph{Poly'19 co-located at VLDB 2019}}.
\newblock


\bibitem[\protect\citeauthoryear{Shah, Banakar, Shastri, Wasserman, and
  Chidambaram}{Shah et~al\mbox{.}}{2019}]%
        {gdpr-storage}
\bibfield{author}{\bibinfo{person}{Aashaka Shah}, \bibinfo{person}{Vinay
  Banakar}, \bibinfo{person}{Supreeth Shastri}, \bibinfo{person}{Melissa
  Wasserman}, {and} \bibinfo{person}{Vijay Chidambaram}.}
  \bibinfo{year}{2019}\natexlab{}.
\newblock \showarticletitle{{Analyzing the Impact of GDPR on Storage Systems}}.
  In \bibinfo{booktitle}{\emph{USENIX HotStorage}}.
\newblock


\bibitem[\protect\citeauthoryear{Shastri, Wasserman, and Chidambaram}{Shastri
  et~al\mbox{.}}{2019}]%
        {gdpr-sins}
\bibfield{author}{\bibinfo{person}{Supreeth Shastri}, \bibinfo{person}{Melissa
  Wasserman}, {and} \bibinfo{person}{Vijay Chidambaram}.}
  \bibinfo{year}{2019}\natexlab{}.
\newblock \showarticletitle{{The Seven Sins of Personal-Data Processing Systems
  under GDPR}}. In \bibinfo{booktitle}{\emph{USENIX HotCloud}}.
\newblock


\bibitem[\protect\citeauthoryear{Solon}{Solon}{2018}]%
        {cambridge-analytica}
\bibfield{author}{\bibinfo{person}{Olivia Solon}.}
  \bibinfo{year}{2018}\natexlab{}.
\newblock \bibinfo{title}{{Facebook says Cambridge Analytica may have gained
  37{M} more users' data}. {I}n \emph{The Guardian}}.
\newblock
  \bibinfo{howpublished}{\url{https://www.theguardian.com/technology/2018/apr/04/facebook-cambridge-analytica-user-data-latest-more-than-thought}}.
\newblock


\bibitem[\protect\citeauthoryear{Sweeney}{Sweeney}{2018}]%
        {usa-today-no-ads}
\bibfield{author}{\bibinfo{person}{Erica Sweeney}.}
  \bibinfo{year}{2018}\natexlab{}.
\newblock \bibinfo{title}{{Many publishers' EU sites are faster and ad-free
  under GDPR}. {I}n \emph{Marketing Dive}}.
\newblock
  \bibinfo{howpublished}{\url{https://www.marketingdive.com/news/study-many-publishers-eu-sites-are-faster-and-ad-free-under-gdpr/524844/}}.
\newblock


\bibitem[\protect\citeauthoryear{Targett}{Targett}{2018}]%
        {pre-gdpr-numbers}
\bibfield{author}{\bibinfo{person}{Ed Targett}.}
  \bibinfo{year}{2018}\natexlab{}.
\newblock \bibinfo{title}{{6 Months, 945 Data Breaches, 4.5 Billion Records.}
  {I}n \emph{Computer Business Review}}.
\newblock
  \bibinfo{howpublished}{\url{https://www.cbronline.com/news/global-data-breaches-2018}}.
\newblock


\bibitem[\protect\citeauthoryear{Tesfay, Hofmann, Nakamura, Kiyomoto, and
  Serna}{Tesfay et~al\mbox{.}}{2018}]%
        {gdpr-www}
\bibfield{author}{\bibinfo{person}{Welderufael Tesfay}, \bibinfo{person}{Peter
  Hofmann}, \bibinfo{person}{Toru Nakamura}, \bibinfo{person}{Shinsaku
  Kiyomoto}, {and} \bibinfo{person}{Jetzabel Serna}.}
  \bibinfo{year}{2018}\natexlab{}.
\newblock \showarticletitle{I Read but Don't Agree: Privacy Policy Benchmarking
  using Machine Learning and the EU GDPR}. In
  \bibinfo{booktitle}{\emph{Companion Proceedings of the The Web Conference
  2018}}. \bibinfo{pages}{163--166}.
\newblock


\bibitem[\protect\citeauthoryear{Utz, Degeling, Fahl, Schaub, and Holz}{Utz
  et~al\mbox{.}}{2019}]%
        {uninformed-consent}
\bibfield{author}{\bibinfo{person}{Christine Utz}, \bibinfo{person}{Martin
  Degeling}, \bibinfo{person}{Sascha Fahl}, \bibinfo{person}{Florian Schaub},
  {and} \bibinfo{person}{Thorsten Holz}.} \bibinfo{year}{2019}\natexlab{}.
\newblock \showarticletitle{(Un) informed Consent: Studying GDPR Consent
  Notices in the Field}. In \bibinfo{booktitle}{\emph{ACM CCS}}.
\newblock


\bibitem[\protect\citeauthoryear{Vardi}{Vardi}{2018}]%
        {move-fast-break-things}
\bibfield{author}{\bibinfo{person}{Moshe Vardi}.}
  \bibinfo{year}{2018}\natexlab{}.
\newblock \showarticletitle{{Move Fast and Break Things}}.
\newblock \bibinfo{journal}{\emph{Commun. ACM}} \bibinfo{volume}{61},
  \bibinfo{number}{9} (\bibinfo{year}{2018}).
\newblock


\bibitem[\protect\citeauthoryear{Wachter, Mittelstadt, and Floridi}{Wachter
  et~al\mbox{.}}{2017a}]%
        {gdpr-no-right-to-explanation}
\bibfield{author}{\bibinfo{person}{Sandra Wachter}, \bibinfo{person}{Brent
  Mittelstadt}, {and} \bibinfo{person}{Luciano Floridi}.}
  \bibinfo{year}{2017}\natexlab{a}.
\newblock \showarticletitle{Why a right to explanation of automated
  decision-making does not exist in the general data protection regulation}.
\newblock \bibinfo{journal}{\emph{International Data Privacy Law}}
  \bibinfo{volume}{7}, \bibinfo{number}{2} (\bibinfo{year}{2017}),
  \bibinfo{pages}{76--99}.
\newblock


\bibitem[\protect\citeauthoryear{Wachter, Mittelstadt, and Russell}{Wachter
  et~al\mbox{.}}{2017b}]%
        {counterfactual}
\bibfield{author}{\bibinfo{person}{Sandra Wachter}, \bibinfo{person}{Brent
  Mittelstadt}, {and} \bibinfo{person}{Chris Russell}.}
  \bibinfo{year}{2017}\natexlab{b}.
\newblock \showarticletitle{Counterfactual explanations without opening the
  black box: Automated decisions and the GDPR.(2017)}.
\newblock \bibinfo{journal}{\emph{Harvard Journal of Law \& Technology}}
  \bibinfo{volume}{31} (\bibinfo{year}{2017}), \bibinfo{pages}{841}.
\newblock


\end{thebibliography}

\end{document}